\documentclass[10pt,sigconf,authorversion,nonacm]{cidr-2027}
\usepackage{graphicx} %
\usepackage{tabularray}
\usepackage{xspace}
\usepackage{multirow}
\usepackage{subcaption}
\usepackage{enumitem}

\newcounter{packednmbr}

\newenvironment{packedenumerate}{\begin{list}{\thepackednmbr.}{\usecounter{packednmbr}\setlength{\itemsep}{0.5pt}\addtolength{\labelwidth}{-14pt}\setlength{\leftmargin}{\labelwidth}\setlength{\listparindent}{\parindent}\setlength{\parsep}{1pt}\setlength{\topsep}{0pt}}}{\end{list}}

\newcounter{qacount}

\newif\ifcomments
\commentsfalse

\ifcomments
\newcommand{\todo}[1]{\textcolor{red}{#1}}
\newcommand{\vyas}[1]{{\footnotesize\color{red}[VS: #1]}}
\newcommand{\milind}[1]{{\footnotesize\color{purple}[MS: #1]}}
\newcommand{\zeying}[1]{{\footnotesize\color{brown}[ZZ: #1]}}
\newcommand{\alan}[1]{{\footnotesize\color{orange}[Alan: #1]}}
\newcommand{\lesley}[1]{{\footnotesize\color{blue}[lesley: #1]}}

\else
\newcommand\todo[1]{}
\newcommand{\vyas}[1]{}
\newcommand{\milind}[1]{}
\newcommand{\zeying}[1]{}
\newcommand{\alan}[1]{}
\newcommand{\lesley}[1]{}

\fi

\newcommand{\mypara}[1]{\smallskip\noindent{\bf {#1}:}~}

\newcommand\sysname{\texttt{ASAP}\xspace}
\newcommand\sysnameFullForm{Application Semantic-Aware Processing\xspace}

\newcommand\olly{observability\xspace}

\newcommand\uc{application\xspace}
\newcommand\ucs{applications\xspace}
\newcommand\Uc{Application\xspace}

\newcommand\primitive{primitive\xspace}
\newcommand\primitives{primitives\xspace}

\newcommand\Primitives{Primitives\xspace}

\newcommand\cpl{control plane\xspace}

\newcommand\dpl{data plane\xspace}

\newcommand\CSP{CSP\xspace}
\newcommand\CSPlong{Cost-Scale-Performance\xspace}
\newcommand\CTSA{CTSA\xspace}
\newcommand\CTSAlong{Collect-Transmit-Store-Analyze\xspace}

\newcommand\lifecycle{lifecycle\xspace}
\newcommand\Lifecycle{Lifecycle\xspace}

\newcommand\ffttView{forest for the trees\xspace}

\newcommand\PromSketch{PromSketch\xspace}
\newcommand\microMon{\textmu Mon\xspace}
\newcommand\microView{\textmu View\xspace}
\newcommand\ASAPQuery{ASAPQuery}

\newcommand\GenT{T-PACK}

\newcommand\Deepcap{DeePCAP}

\title{\sysname: Reimagining the Data \Lifecycle using \underline{A}pplication \underline{S}emantic-\underline{A}ware \underline{P}rocessing }

\author{
Milind Srivastava\textsuperscript{1},
Zeying Zhu\textsuperscript{2},
Yajie Zhou\textsuperscript{2},
Yancheng Yuan\textsuperscript{2},
Fenghao Dong\textsuperscript{1},
Peilin Xin\textsuperscript{1},
Zaoxing Liu\textsuperscript{2},
Vyas Sekar\textsuperscript{1}
\\
\textsuperscript{1}Carnegie Mellon University
\quad
\textsuperscript{2}University of Maryland \\ 
}

\renewcommand{\shortauthors}{
Milind Srivastava,
Zeying Zhu,
Yajie Zhou,
Yancheng Yuan,
Fenghao Dong,
Peilin Xin,
Zaoxing Liu,
Vyas Sekar}

\begin{document}

\begin{abstract}

Across many domains (e.g., observability, networking, security),  data processing pipelines face what we refer to as  the {\em \CSP problem}: achieving low \underline{C}ost at large \underline{S}cale, while maintaining high \underline{P}erformance. 
In response, we see  several efforts  to tackle  \CSP in various stages of the \CTSAlong data \lifecycle; such as approximate query processing in databases   or  sketches in network  routers. 
Our work is driven by the simple insight: ``seeing the \ffttView''. These proposed solutions  (e.g., AQP, sketching, compression, rollup) addressing \CSP share a common property --- they exploit semantic-preserving  opportunities to support \uc  needs.
In this paper, we make a case for  \sysname, a paradigm that makes \sysnameFullForm (\sysname) a first-class design principle in data processing pipelines. We argue that by taking a {\em unified} view {\em across \sysname \primitives} developed in different domains, {\em across the entire data \lifecycle}, we can  unlock new opportunities  to tackle the \CSP problem. 
 In particular, we can: (i)  enable novel cross-\lifecycle optimizations such as  analytics run directly on  sketches computed at the source; (ii) leverage \primitives developed in other \uc domains; and  (iii) enable widespread adoption of these powerful techniques.
We discuss research challenges in socializing the benefits of the \sysname paradigm, and show preliminary evidence that adopting  \sysname    can yield up to 3 orders of magnitude  improvements in the \CSP tradeoff for many \uc domains.

\end{abstract}

\maketitle

\section{Introduction}

Modern systems such as cloud infrastructure, networks, and enterprise software increasingly depend on data for various operational tasks: from incident detection~\cite{li2020gandalf} and root cause analysis~\cite{gan2021sage} to online control~\cite{rzadca2020autopilot} and long-term planning~\cite{ eriksen2023global}. 
As systems grow in size and complexity, they face a common set of pressures that we refer to as the  \CSPlong (\CSP) problem: 
\begin{packedenumerate}
\item 
\textbf{Cost:} 
Resource usage and dollar costs of data-driven tasks have skyrocketed.
For example, more than a third of enterprises spend over \$1M annually on cloud observability~\cite{logicmonitor2025obsbudget}, and costs are growing 40--48\% year over year~\cite{grafana2025obssurvey, honeycomb2025obscost}. 
The cost problem is exacerbated by rising memory prices~\cite{memory-prices-1} and energy demands~\cite{energy-demand-1}.

\item
\textbf{Scale:} Data volume is growing because deployments are larger and more complex, and because operators require finer-grained signal data to understand these systems~\cite{dean2013tail}. Critically, this growth is no longer just human-scale. Machine-generated traffic, from bots and AI agents~\cite{ai_cost_at_scale, cnbc2026aibots}, now makes up the majority of the activity that operational systems must observe and secure~\cite{liu2025supporting}. 

\item
\textbf{Performance:} The performance expectations of data consumers are rising. Humans expect interactive dashboards and fast troubleshooting, while AI agents increasingly sit in the loop, and require timely signals for closed-loop decisions that compound the scale problem with tighter latency and higher throughput requirements.

\end{packedenumerate}

In response to the \CSP problem, researchers (including  the  authors) and practitioners  from various communities have proposed numerous solutions.  
 For example, in  network monitoring, research changes the   \textit{collection} stage  to  use   sketches  in programmable router hardware~\cite{liu2016univmon}.
 Similarly,  time-series databases in various \ucs apply  fidelity-reducing roll-ups to reduce \textit{analytics} latency and \textit{storage} costs~\cite{thanos, prometheus}.   Approximate query processing (AQP)  uses sampling techniques  to reduce \textit{analytics} cost and latency~\cite{verdictdb, pilotdb, blinkdb}.
 Other  ML-based generative compression approaches  reduce \textit{transmission} and \textit{storage} costs of high-dimensional telemetry~\cite{zhou2026prvtel, deepcap-workshop}.

While at first glance, these efforts (including our own) seem unrelated, we observe that they share a common theme: ``preserving the information needed to satisfy \uc semantics, rather than preserving all the raw data''.
Various ``\textit{\primitives}'' like sampling, sketching, roll-ups and compression are all manifestations of this idea. We refer to these as \textit{\sysnameFullForm (\sysname) \primitives}.

\begin{figure}[t]
    \centering
    \includegraphics[width=\linewidth]{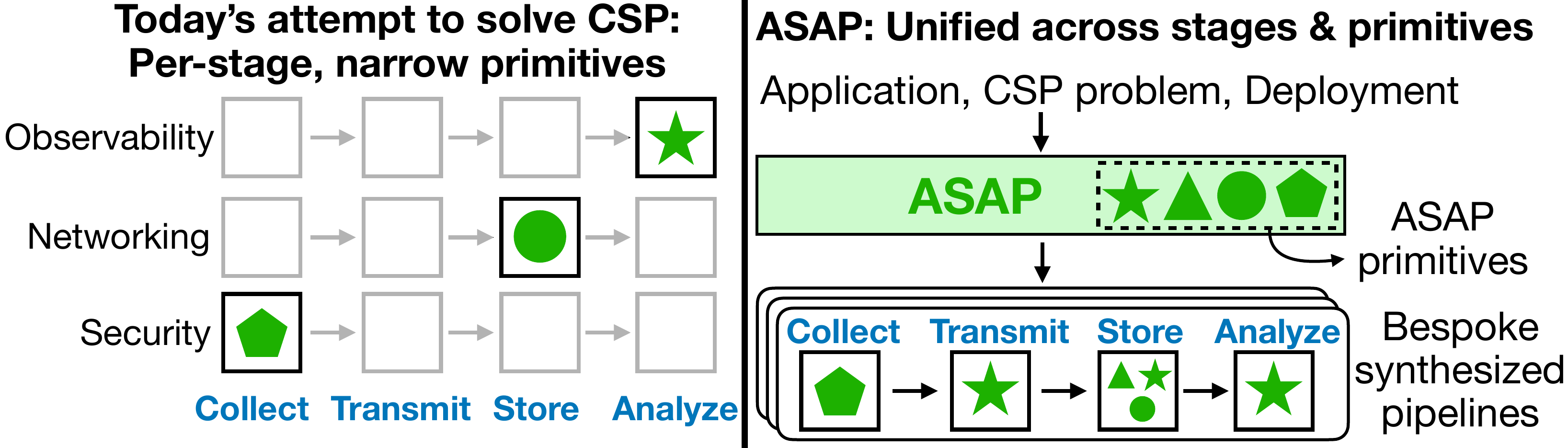}
    \vspace{-0.6cm}
    \caption{\sysname makes \underline{A}pplication \underline{S}emantic-\underline{A}ware \underline{P}rocessing \primitives first-class citizens to enable cross-domain and cross-\lifecycle benefits.}
    \label{fig:intro}
\end{figure}

Building on this, we propose the \sysname paradigm (Figure~\ref{fig:intro}): making \textit{\sysnameFullForm} (\sysname) a first-class design principle for current and future data processing  pipelines. 
Rather than rediscovering \primitive-based solutions as isolated, domain-specific optimizations at individual \CTSAlong (\CTSA) stages,
we argue for a {\em unified} view across the \CTSA \lifecycle and across \sysname \primitive families, to address \CSP problems more systematically across \ucs and domains.

By explicitly systematizing the \sysname paradigm, we posit that we can enable new  cross-domain, cross-\lifecycle, and cross-\primitive benefits (details in \S\ref{subsec:case_for_unified}). This will allow us to translate \CSP benefits from one domain to another, and enable us to combine optimization opportunities across the \CTSA \lifecycle. Further, we can reduce the barrier to adopt \sysname \primitives via common abstraction layers that allow diverse communities to leverage these powerful techniques.

Note that \sysname is not a ``replacement'' for the impressive body of work on various kinds of information synopses including AQP, sketches, wavelets, and others~\cite{cormode2011synopses}. Rather, we view it as a timely attempt to revisit, unify, and socialize the benefits of work scattered across different domains, \primitives, and stages of the data \lifecycle.
We believe the time is right for ASAP, as strong \uc ``pulls'' are now converging  with the attendant technology ``pushes''.
\CSP pressures have already been growing across domains and \ucs, and the shift from human-generated to machine-generated data only compounds this challenge.
At the same time, we see the maturation of \sysname \primitives  in \ucs~\cite{prometheus, bigquery-approx-agg, blinkdb, redis2022countminsketch, ddsketch}, and advances in AI-assisted system design~\cite{adrs, selfdefining, glia, grayeli2024lasr} that enable rapid prototyping and design space exploration.

To help operators fully realize cross-domain, cross-\lifecycle, and cross-\primitive benefits, we envision the \sysname framework as shown in the right half of Figure~\ref{fig:intro}.
The operator specifies a problem spec with their \ucs and \CSP bottlenecks, and a deployment spec describing their deployment, constraints and historical data and queries. Using a library of \primitives, the \sysname framework automatically synthesizes a bespoke \sysname-based \CTSA pipeline that meets \uc fidelity requirements and tackles \CSP bottlenecks.

As early evidence of \sysname's promise, we designed and deployed \sysname-enabled pipelines for three domains (\S\ref{sec:early_promise}): metrics \olly, distributed tracing, and network flow monitoring. Optimizing across domains and \lifecycle stages with \sysname \primitives, we observe up to 3 orders of magnitude benefits in \CSP tradeoffs over baselines. 
For e.g., in a metrics observability \uc, \sysname reduces transmission, and analysis costs by 48$\times$ and 81$\times$ , while reducing memory usage by up-to 7000$\times$ and accelerating queries by up-to 3000$\times$.

We identify key research challenges in fully realizing the \sysname vision and its benefits.
Synthesizing a pipeline requires selecting \primitives, placing and configuring them across \lifecycle stages, and defining how they interoperate. Operating a pipeline requires dynamic reconfiguration of \primitives as workloads shift and supporting complex queries beyond simple aggregates. Easy deployment requires incremental adoption into existing tech stacks.

\section{Status quo: Isolated efforts}
\label{sec:motivation}

\begin{table*}[]
\footnotesize
\centering
\begin{tabular}{|l|l|l|l|l|}
\hline
\textbf{Domain} &
  \textbf{\Uc} &
  \textbf{\Lifecycle Stage} &
  \textbf{Solutions} &
  \textbf{\Primitives used} \\ \hline
\multirow{3}{*}{Observability} &
  Fine-grained telemetry monitoring &
  Collection &
  \microView~\cite{cornacchiaobservability} &
  Sketches \\ \cline{2-5}
 & 
  Real-time metrics dashboard &
  Analytics &
  \begin{tabular}[c]{@{}l@{}}PromSketch~\cite{zhu2025approximation}, \ASAPQuery~\cite{asapquery}\end{tabular} &
  Sketches \\ \cline{2-5} 
 &
  Real-time trace dashboard &
  Transmission &
  \GenT~\cite{gent-nsdi} &
  \begin{tabular}[c]{@{}l@{}}Statistical compression, \\ Graph summarization\end{tabular} \\ \cline{2-5}
  &
  Real-time trace dashboard &
  Transmission &
  Hindsight~\cite{hindsight} &
  \begin{tabular}[c]{@{}l@{}}Sampling\end{tabular} \\ \hline
 
\multirow{2}{*}{Networking} &
  Longitudinal PCAP analytics &
  Storage &
  \Deepcap~\cite{deepcap-workshop} &
  \begin{tabular}[c]{@{}l@{}}Deep generative models, Wavelets\end{tabular} \\ \cline{2-5} 
 &
  Fine-grained telemetry monitoring &
  Collection &
  \microMon~\cite{micromon} &
  Sketches, Wavelets \\ \hline
\multirow{2}{*}{Security} &
  Intrusion detection &
  Analytics &
  Count-Less~\cite{kim2023robust} &
  Sketches \\ \cline{2-5} 
 &
  Causality analysis &
  Storage &
  SEAL~\cite{seal}, NodeMerge~\cite{nodemerge} &
  Graph summarization \\ \hline
\multirow{2}{*}{Business analytics} &
  Decision support &
  Analytics &
  \begin{tabular}[c]{@{}l@{}}BlinkDB~\cite{blinkdb},  VerdictDB~\cite{verdictdb}, PilotDB~\cite{pilotdb}\end{tabular} &
  Sampling \\ \cline{2-5} 
 &
  Alternative history analysis &
  Storage, Analytics &
  AHA~\cite{aha} &
  Rollups \\ \hline
\end{tabular}
\caption{Across domains, researchers have leveraged different \primitives to tackle \CSP problems for specific \ucs
, and stages of the \CTSA \lifecycle.}
\label{tab:usecases}
\vspace{-0.5cm}
\end{table*}

We begin by discussing the \CSP problem in different domains. We highlight specific techniques (e.g., sampling, sketching, wavelets, rollups)  that experts in these domains have used for specific stages of the \CTSAlong data \lifecycle (Table~\ref{tab:usecases}). 
We then highlight some missed opportunities that motivate the \sysname paradigm.

\mypara{Observability}
Cloud observability relies on metrics, logs, and traces for alerting, debugging, and capacity planning. 
High-cardinality metrics, fine-grained telemetry, and distributed traces raise collection, transmission, storage, and query costs. 
Existing systems try to optimize for a specific \CTSA stage.
For example, \microView~\cite{cornacchiaobservability} uses sketches for fine-grained telemetry \textit{collection}, while \PromSketch~\cite{zhu2025approximation} and \ASAPQuery~\cite{asapquery} use them to reduce \textit{analytics} cost. \GenT~\cite{gent-nsdi} and HindSight~\cite{hindsight} use statistical compression and sampling, respectively, to reduce the \textit{transmission} cost.

\mypara{Networking}
Network monitoring \ucs collect packet and flow-level telemetry from switches and hosts, and analyze it for troubleshooting, control decisions, and longitudinal traffic analysis.
Here, the \CSP bottleneck depends on the timescale.
Short-lived network events require costly fine-grained data collection, while longitudinal analysis makes raw data retention costly. 
\microMon~\cite{micromon} uses sketches and wavelets to optimize \textit{collection} stage cost and performance.
\Deepcap~\cite{deepcap-workshop} uses wavelets with deep generative compression to reduce the \textit{storage} cost of long-term packet data retention.

\mypara{Security}
Security \ucs use network traffic, logs, and system events, for causality analysis and intrusion detection.
As intrusion detection demands high analytics throughput at low cost, 
prior work uses sketches to optimize the \textit{analytics} stage~\cite{kim2023robust}.
Causality analysis incurs storage costs from retaining large provenance graphs. SEAL~\cite{seal} and NodeMerge~\cite{nodemerge} use graph summarization to reduce \textit{storage} cost.

\mypara{Business analytics}
Business analytics \ucs collect product and user records, transmit them to analytical backends, store historical datasets, and analyze them for decision support, regression analysis, and ``what-if'' exploration.
Here, the \CSP bottleneck often lies in storage and analytics, as retaining and scanning raw historical data is costly.
Approximate Query Processing (AQP) systems~\cite{blinkdb,verdictdb,pilotdb} use sampling to reduce \textit{analytics} latency and resource cost for decision-support workloads.
AHA~\cite{aha} uses hierarchical rollups to reduce both \textit{storage} and \textit{analytics} cost for longitudinal time-series data analysis.

\mypara{Missed opportunities}
While numerous solutions have been designed, we observe that these are largely siloed: each targets a specific domain, a narrow \uc, and a specific \CTSA \lifecycle stage. This   leaves optimization opportunities untapped across domains, \ucs, and \lifecycle stages. 
For instance, instead of using statistical compression to reduce the \textit{transmission} cost of metrics \olly, couldn't we also use it at the backend, to tackle \textit{storage} and \textit{analytics} bottlenecks? 
What if instead of using wavelets only for network \textit{storage} bottlenecks, we also use them to improve \textit{analytics} performance?  
What if we could combine sampling, sketches, and wavelets into a single pipeline to boost the performance of an \textit{analytics} workload that cannot be tackled by any single one of them?

\section{\sysname Vision and Overview}
To unlock these missed opportunities, we first identify the common structure underlying existing solutions, and then use it to motivate the \sysname paradigm.

\subsection{The case for a unified view}
\label{subsec:case_for_unified}

At first glance, a sketch in a router, a wavelet transform for packet traces, and sampling for database queries may seem unrelated. Yet, they share a common idea: each of them is built to satisfy a specific type of \textit{\uc semantics}. \Uc semantics refer to the properties of raw data that an \uc cares about. For e.g., a business intelligence application may care about aggregates like quantiles, while a networking dashboard may care about temporal behavior like inter-arrival packet time.

In essence, each  solution from  \S\ref{sec:motivation} applies an \textit{\uc semantic-aware \primitive}: it processes raw data and   only preserves the properties that the \uc needs. From this perspective, \primitive families are defined by the semantics they preserve, not by the domains in which they were invented. Sampling provides general-purpose summaries; sketches preserve aggregates; histograms preserve distributions; wavelets preserve spatial or temporal frequency structure; rollups preserve hierarchies; coresets~\cite{feldman2012data} %
preserve the structure needed for specific optimizations; and generative models capture high-dimensional relationships.

This common abstraction of \textit{\uc semantic-aware \primitives} suggests a better way to build \CSP solutions.
Instead of re-implementing a \primitive inside a single \CTSA stage of a domain-specific system, we can build a shared abstraction over \sysname \primitives, select one or more of them based on \uc semantic needs, and deploy them across multiple stages.
We argue that we can unlock  substantial benefits by taking this unified view across domains and \ucs, \lifecycle stages, and \primitives.

\mypara{Reusing \primitives across domains} A useful \primitive should not remain trapped in the domain where it was invented. For example, UnivMon~\cite{liu2016univmon} uses a single sketch to support several network-monitoring queries. Database and observability systems ask many of the same question (top-$k$, cardinality, and entropy over high-volume streams) and can therefore benefit from UnivMon. A unified view allows reuse of not only the \primitive implementation, but also its fidelity guarantees and configuration logic; thus reducing the barrier to adopt \primitives. For example, the logic for sizing a sketch to meet a target error can transfer from network monitoring to observability metrics. 
The same opportunity applies to other \primitives. Wavelets, used widely for image compression and then applied to long-term packet retention~\cite{deepcap-workshop}, could also help store time-series data in other domains.
The unified view is not that one \primitive fits every application, but that a successful idea should become a shared building block across domains and \ucs.

\mypara{Evolving \primitives as requirements change}
A pipeline should not be locked into a fixed \uc--\primitive pairing. The workload may evolve, or a better \primitive may become available. In either case, treating the \primitive as a replaceable component lets the system adapt without rebuilding the surrounding pipeline. Consider a security analytics \uc that initially uses sketches to detect heavy hitters and volumetric attacks. If the threat shifts to periodic beaconing or short bursts, temporal structure becomes more important than aggregate counts, and the system can replace the sketch with a wavelet representation. Alternatively, if a more accurate or efficient sketch is developed, the system can swap it in while supporting the same queries. 

\mypara{Combining \primitives across the data \lifecycle} The benefits can be magnified if we apply the \sysname insight across the \lifecycle. For example, PromSketch~\cite{zhu2025approximation} and ASAPQuery~\cite{asapquery} use sketches to accelerate analysis, but the raw data has already been collected and transmitted by then. Instead, if a metrics observability collector can build the sketch at the data source, transmit the compact summary, and the analytics backend can answer queries directly using the summary, the benefits  magnify  across stages. Different semantic-preserving \primitives can also be stacked when an \uc workload requires multiple properties. For NetFlow analysis, sketches serve real-time aggregate queries, wavelets preserve time-varying behavior of each flow, and deep generative compression captures redundancy across many flows. These components address different stages of the same workload: \textit{storage} and \textit{analysis}.

\subsection{The \sysname framework}
Motivated by these benefits, we envision \sysname, a new data processing paradigm that argues for  a unified view of \textit{\uc semantic-aware \primitives}, and treats them as first-class citizens to tackle diverse \CSP problems. 
\sysname  reflects a shift from \textit{data-centric} to \textit{information-centric} pipelines; i.e.,  instead of treating raw data as the  atomic unit to collect, transmit, store, and analyze, \sysname treats the \textit{information} the \uc requires as the basic unit.

To bring this paradigm to life, we envision  the \sysname framework  as shown in the right half of Figure~\ref{fig:intro}. 
The operator specifies a problem specification that describes the \uc of interest and \CSP bottlenecks, and a deployment specification that describes the current data \lifecycle pipeline, deployment constraints (e.g.  compatibility with existing technology stacks, resource budgets), and historical data/query workloads.
Using a library of \sysname \primitives, the \sysname framework automatically synthesizes a bespoke \sysname-based \CTSA pipeline that respects deployment constraints. This pipeline consists of a \dpl and a \cpl.
The \dpl consists of \CTSA stages, computes \primitives from raw data, transmits and stores primitive instances, and serves \uc queries from them.
The \cpl continuously monitors the \dpl's data and query workload dynamics, resource utilization, and \uc fidelity, and reconfigures \primitives at runtime, deploys new \primitive instances, or reallocates resources between \primitives, as needed.

For example, in a metrics observability pipeline, an operator may need to support tail-latency and heavy-hitter queries while facing high network and storage costs. The deployment specification may further require compatibility with an existing Prometheus backend and impose a strict CPU budget at collectors. Based on these specifications, the \sysname framework may select quantile and frequency-estimation \primitives, place their computation at the collectors, store their compact instances in the existing backend, and synthesize the corresponding query-processing logic.

\section{Early Promise of \sysname}
\label{sec:early_promise}

\begin{figure}[t]
    \centering
    \includegraphics[width=\linewidth]{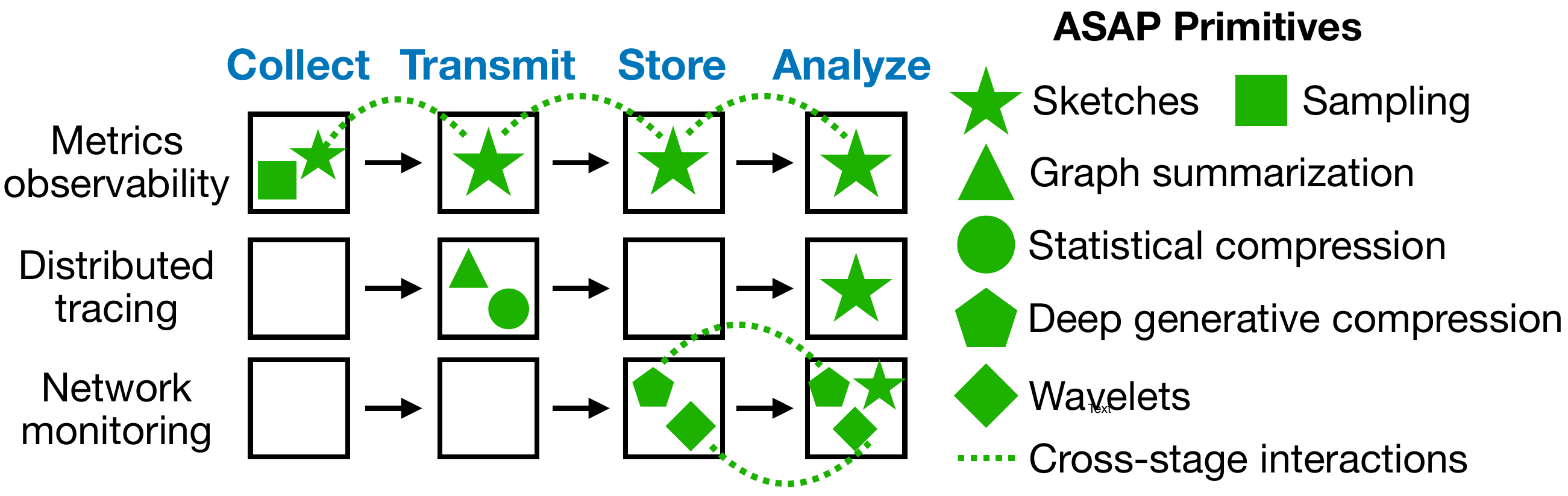}
    \vspace{-0.8cm}
    \caption{Three implemented \sysname-based pipelines to show early promise of the \sysname paradigm.}
    \label{fig:case_studies}
    \vspace{-0.25cm}
\end{figure}

To demonstrate \sysname's potential, we design and implement \sysname-enabled pipelines for three domains, across multiple \lifecycle stages and \primitive combinations (Figure~\ref{fig:case_studies}).

\begin{figure}[t]
    \centering
    \includegraphics[width=\linewidth]{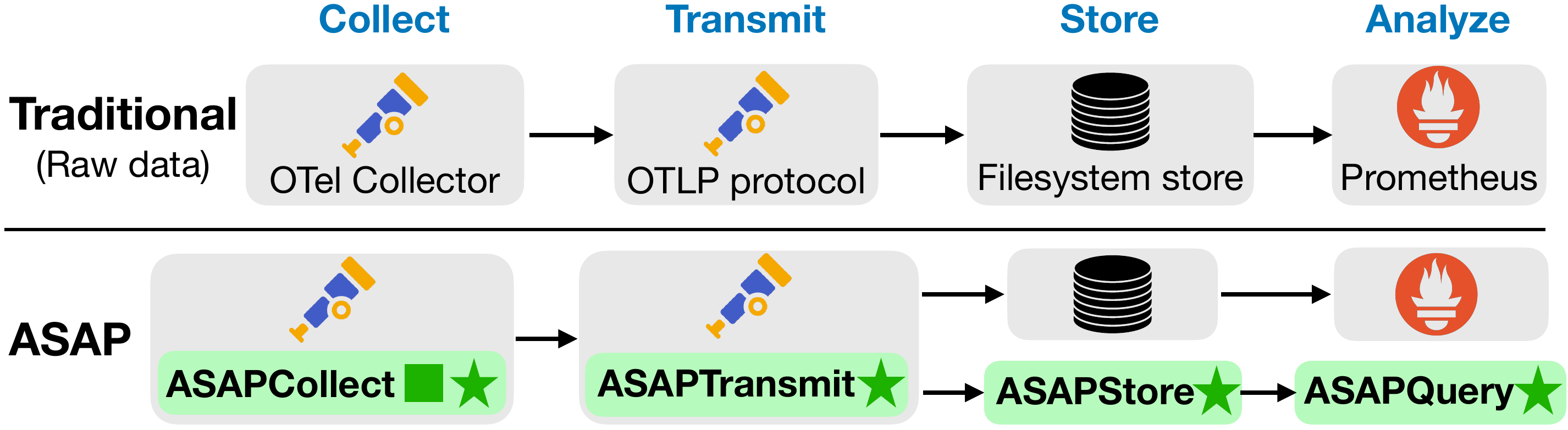}
    \vspace{-0.8cm}
    \caption{\sysname pipeline for metrics \olly.}
\label{fig:case_studies_metrics_o11y}
\end{figure}

\mypara{Metrics \olly}
Real-time metrics dashboards periodically issue aggregate queries over recently ingested data, like \texttt{quantile}, \texttt{topk}, and \texttt{cardinality}. The top half of Figure~\ref{fig:case_studies_metrics_o11y} illustrates a traditional pipeline for serving these. The OTel~\cite{otel} Collector records raw metrics and transmits them via the OTel Protocol (OTLP) to Prometheus, which stores the data on the local filesystem. Prometheus's query engine executes queries over these metrics to serve dashboards and alerts.
The pipeline operator seeks to reduce the total cost of ownership (TCO). They are willing to modify the collector, but require Prometheus to be completely unmodified and to continue storing raw metrics.

Seeing a parallel between the metrics dashboards \uc and switch-based network monitoring, where sketches reduce the cost of flow monitoring~\cite{liu2016univmon}, we build an \sysname pipeline that applies sketches at each stage of the \CTSA \lifecycle (Figure~\ref{fig:case_studies_metrics_o11y}). The pipeline reduces TCO while satisfying the deployment constraints.
ASAPCollect~\cite{asapcollector_repo} inserts raw metrics into sketches, that are then transported over OTLP by ASAPTransmit~\cite{asapcollector_repo} to ASAPStore~\cite{asapquery}. ASAPStore indexes these sketches, and provides them to ASAPQuery~\cite{asapquery} that implements a sketch-based query engine and serves metrics dashboards using the received sketches.
To satisfy the deployment constraints, ASAPStore and ASAPQuery leave Prometheus unmodified, and operate as a ``drop-in'' middlebox between Prometheus and the dashboard. Time-series metrics that are not incorporated into sketches continue to be ingested and stored as raw data in Prometheus.

We evaluate this pipeline on a stream of 1M time-series generated by 10 hosts (each generating 100k time-series). Data is ingested every 100ms (10 samples/series/sec), for a resulting throughput of 10M data points/sec. 
We run \texttt{quantile}, \texttt{cardinality}, and \texttt{top-k} queries ``grouped by'' each host, with a 95\% relative accuracy target.
We compare our pipeline to Prometheus~\cite{prometheus} and observe 28-3000$\times$ query performance improvement depending on the exact query.
To get an upper bound of this pipeline's benefits at hyper-scale (100M time-series), we linearly extrapolate resource usage by 100$\times$, and obtain 48$\times$ and 81$\times$ reduction in transmission and analytics cost, respectively. ASAPCollect increases the collection CPU usage due to sketch ingestion by up-to 2$\times$, but reduces memory usage by 49--7000$\times$.
\footnote{We leave a more detailed evaluation to future work.}

\mypara{Distributed tracing}
Real-time trace dashboards analyze microservice traces for  aggregates (e.g., \texttt{quantile}/\texttt{topk}) and ``RED'' metrics~\cite{traces_red}: rate, error, and duration. Transmitting all raw traces from the collector to the backend is a bottleneck. 
Statistical compression and graph summarization can model the high-dimensional and graphical nature of traces, required by RED queries.
We implement an \sysname pipeline that uses statistical compression and graph summarization at the \textit{transmit} stage, and sketches at the \textit{analyze} stage (using ASAPQuery~\cite{asapquery}), and evaluate it on a public Uber dataset~\cite{uber_data}
\footnote{The dataset contains 531K traces (746M spans) spanning multiple days, with randomized start times. To simulate realistic data, we sample 50K traces and remap them into a 1-second epoch, yielding \textasciitilde 800k spans.}.
We achieve a 140$\times$ compression ratio, with 50$\times$ lower \textit{transmission} cost (while matching the fidelity of 1:3 sampling) and \textasciitilde5$\times$ lower \textit{analysis} cost ($\geq$ 95\% relative accuracy).

\mypara{Network monitoring}
Packet and flow (e.g., Cisco Netflow) data is used for real-time aggregate dashboards and longitudinal tasks like forecasting.
We notice that wavelets (to preserve time-varying behavior) and deep generative compression (to compress across flows) together reduce long-term storage costs.
We evaluate an \sysname pipeline with these \primitives at \textit{storage}, and sketches at \textit{analyze} on the CAIDA dataset~\cite{caida}. We observe 600$\times$ storage cost reduction with over 50\% fidelity improvement compared with the state-of-the-art~\cite{netshare}, and 4--10$\times$ reduction in query latency compared to Clickhouse~\cite{clickhouse} ($\geq$ 95\% relative accuracy).

\section{Open Challenges}

We discuss open challenges and opportunities for new research enablers for fully realizing \sysname's unified vision across domains, \ucs, and \lifecycle stages.

\mypara{Selecting \primitives for \ucs}
Selecting suitable \primitives for \ucs requires translating an \uc's semantics, fidelity requirements, cost and performance constraints, and input-data characteristics into concrete requirements. The \sysname framework must  match these requirements against the capabilities and tradeoffs of candidate \primitives. When no existing \primitive provides the required capabilities, this process should also reveal opportunities for composing existing \primitives or designing new ones.

\mypara{Cross-\lifecycle \primitive placement \& configuration}
In our prototypes, we manually placed and configured \primitives across the \lifecycle. Automating this requires end-to-end optimization of fidelity, cost, and performance while ensuring that upstream reductions preserve the information required downstream. For example, sketches and wavelets can both reduce transmission cost, but a sketch configured for aggregate queries may not support point queries or finer-grained analysis. \sysname must also balance competing objectives: for example, when analytics fidelity is the primary objective, retaining more wavelet coefficients may improve \uc fidelity with higher transmission cost, and \sysname should keep the cost under the available budget.

\mypara{Cross-\lifecycle \primitive interoperability} 
Once selected, \primitive instances need to be transmitted, stored,
merged, composed, and reconfigured across stages. 
To avoid domain-specific reinvention, \sysname needs standardized self-describing representations, hardware-accelerated implementations (e.g. on GPUs and FPGAs), and expressive APIs that expose capabilities, constraints, composition rules, and fidelity-tuning mechanisms.

\mypara{Dynamic pipeline reconfiguration}
Our pipelines were configured once at startup based on a priori information of the workload structure. In production, this information may be unavailable or workload structure might change after deployment. Understanding how/when to reconfigure \sysname \primitives and pipelines, and possibly using \sysname itself to inform these policies, is an open challenge.

\mypara{Supporting complex queries}
Our prototypes focus on aggregate queries that map naturally
to selected \primitives. Production workloads may contain joins,
nested queries/CTEs, and other complex structures. Supporting
these requires determining how multiple \primitives compose,
how their guarantees propagate through query plans, and when parts of a
query must fall back to exact computation.

\mypara{Incremental adoption}
In our prototypes, we manually chose whether to modify selected stages of an existing pipeline or introduce a drop-in layer. Real deployments constrain which components can be changed, where \primitives can be introduced, and which interfaces must be preserved. The \sysname framework must account for deployment constraints and produce pipelines that support incremental migration rather than requiring a flag-day replacement.

\section{Discussion and Conclusion}
\label{sec:discussion}

\mypara{\sysname in a world of AI coding agents}
Implementing pipelines is faster and cheaper, but coding agents cannot determine the information a pipeline must preserve, or the \sysname \primitives to use and compose.
Rather, easy code generation complements ASAP and, together with AI-assisted 
exploration techniques~\cite{bespoke-olap}, helps synthesize bespoke \CTSA pipelines.

\mypara{\sysname in a world of agentic data consumers}
Agentic consumers strengthen the case for \sysname as they are more sensitive to performance (e.g., throughput, latency)~\cite{ghotikar2026scalingagents, liu2025supporting} and more tolerant of approximation~\cite{liu2025supporting}. 

\mypara{Communicating \primitive guarantees} %
Agentic consumers need error bounds, confidence information, and provenance to incorporate into downstream decisions.
Humans require presentations that make the same guarantees understandable and actionable~\cite{aqp-visualization}. Although users already accept approximation in many settings, they may not recognize when or how it arises~\cite{cormode2011synopses}. 
Thus, communicating uncertainty effectively is important to socialize \sysname's benefits.

\mypara{Energy and carbon efficiency}
Rising energy demand from data processing makes reducing unnecessary computation, movement, and storage increasingly important for limiting energy use and carbon emissions~\cite{hotcarbon}. \sysname is well suited to this goal as it preserves and processes only the information downstream \ucs need. A promising extension is to minimize energy and carbon footprint when synthesizing and configuring \sysname-enabled pipelines.

\mypara{Conclusion}
In this position paper, we identify a common principle behind \primitives developed across domains and stages of the \CTSA \lifecycle for tackling \CSP problems: preserving the information \ucs require rather than building pipelines on raw data. The \sysname paradigm elevates this into a first-class abstraction, enabling \primitives to be reused across domains, composed across capabilities, and carried across \lifecycle stages. Preliminary pipelines show benefits compounding across the data \lifecycle, up-to three orders of magnitude improvements. We believe this information-centric view opens a broad agenda for designing the next generation of efficient data-processing pipelines.

\renewcommand{\bibfont}{\scriptsize}

\bibliographystyle{abbrv} 
\bibliography{ref}

\end{document}